# Quenching of charge and spin degrees of freedom in condensed matter

*Fumitaka Kagawa\*, Hiroshi Oike*

Dr. F. Kagawa, Dr. H. Oike
RIKEN Center for Emergent Matter Science (CEMS), Wako 351-0198, Japan
E-mail: fumitaka.kagawa@riken.jp

**Abstract**

**Electrons in condensed matter have internal degrees of freedom, such as charge, spin and orbital, leading to various forms of ordered states through phase transitions. However, in individual materials, a charge/spin/orbital ordered state of the lowest temperature is normally uniquely determined in terms of the lowest-energy state, i.e., the ground state. Here, we summarize recent results showing that under rapid cooling, this principle does not necessarily hold, and thus, the cooling rate is a control parameter of the lowest-temperature state beyond the framework of the thermo-equilibrium phase diagram. Although the cooling rate utilized in low-temperature experiments is typically $2\times10^{-3}$–$4\times10^{-1}$ K/s, the use of optical/electronic pulses facilitate rapid cooling, such as $10^2$–$10^3$ K/s. Such an unconventionally high cooling rate allows some systems to kinetically avoid a first-order phase transition, resulting in a quenched charge/spin state that differs from the ground state. We also demonstrate that quenched states can be exploited as a non-volatile state variable when designing phase-change memory functions. The present findings suggest that rapid cooling is useful for exploring and controlling the metastable electronic/magnetic state that is potentially hidden behind the ground state.**

# 1. Introduction

The first-order phase transition that occurs during cooling often progresses in a limited temperature range below the transition temperature.[1] This observation suggests that during cooling, the phase-transition kinetics or the phase-transition speed competes with the experimental cooling rate. When the utilized cooling rate is sufficiently slow compared with the phase-transition kinetics, the phase transition will be fully completed when the specimen temperature passes through a certain temperature range in which the first-order transition can progress at the laboratory time scale. In such a "slow-cooling regime," the low-temperature phase realized coincides with the thermo-equilibrium phase diagram. By contrast, when the cooling rate sufficiently exceeds the phase-transition kinetics, the temperature range described is traversed practically before the phase transition starts; consequently, the first-order phase transition is kinetically avoided. Thus, the low-temperature state in such a "quenched regime" is not the same as that in the thermo-equilibrium phase diagram (normally, a supercooled state of the high-temperature phase is realized as a quenched state). Such cooling-rate-dependent behavior may be summarized by noting that a low-temperature state of a first-order-transition system can result in either the most thermodynamically stable phase (the slow-cooling regime) or a metastable quenched state (the quenched regime). Although metastable states have finite lifetimes by definition, they can be unmeasurably long at temperatures well below the transition temperature.

Liquids are one of the most prototypical systems exemplifying the competition between phase-transition kinetics and cooling rates.[1-3] When slowly cooled, liquids crystallize at approximately the melting temperature, $T_m$ (**Figure 1**a). However, this first-order liquid-to-crystal transition is kinetically avoided when rapidly cooled; thus, the liquid can persist as a supercooled liquid, even below $T_m$. Upon further cooling, the thermal motions of constituent atoms or molecules are suppressed and eventually frozen; as a result, a disordered atomic/molecular configuration, called glass or structural glass, is realized as a quenched state. Although kinetic properties are remarkably (and even nontrivially) different between glass and supercooled liquid, no well-defined thermodynamic phase transition exists between them. Therefore, glass may be described as a nontrivial supercooled state of liquids. In this way, depending on the utilized cooling rate, liquids can become either crystals (the slow-cooling regime) or metastable glasses (the quenched regime), as illustrated in Fig. 1b.

An aggregate of carbon atoms also exhibits the competition described above. As seen in the pressure-temperature phase diagram (Fig. 1c),[4] the thermodynamically stable state at room temperature and ambient pressures is graphite, but diamond also exists under ambient conditions as a metastable state. In nature, diamond grows in the Earth's mantle as the thermodynamically stable state in high-temperature and high-pressure environments; then, it is transferred to the surface of the Earth via various pressure-temperature histories. When this travel occurs sufficiently slowly, diamond becomes graphite because the pressure-temperature conditions cross the first-order phase boundary between diamond and graphite. In contrast, when the travel occurs rapidly, for instance, via eruption, the structural transition is kinetically avoided, and consequently, diamond remains as a quenched state, even after the aggregate of carbon atoms reaches ambient conditions. Thus, starting from diamond at high temperatures and high pressures, the ambient-condition state can be either graphite (the slow-cooling regime) or quenched diamond (the quenched regime), depending on the cooling rate (Fig. 1d). We note that a quenched state may be long-range ordered, as in the case of diamond.

Despite the frequent use of rapid cooling in the context of material synthesis to obtain the metastable forms of solid states, rapid cooling has scarcely been exploited to realize metastable electronic/magnetic states in the study of interacting electron/spin systems, probably because it has not been considered capable of dramatically altering the low-temperature state. In fact, as presented below, the highest cooling rate achievable in a conventional cryostat with liquid helium is only 2–4×10$^{-1}$ K/s, and this value is often too low to reach the quenched regimes of interacting electron/spin systems. Therefore, it is unsurprising that the presence of the quenched regime has been experimentally overlooked in many materials.

Here, in this Research News contribution, we highlight recent progress in the application of rapid cooling up to 10$^2$–10$^3$ K/s in (i) strongly correlated electron systems exhibiting first-order charge-ordering transition[5-8] and (ii) a non-centrosymmetric chiral magnet with a first-order transition from a topological spin texture called magnetic skyrmion to a non-topological conical spin texture.[9] In both systems, metastable charge/spin states are realized only when the utilized cooling rate exceeds a threshold cooling rate, which is material dependent. We also show that a charge/spin state can be switched deterministically and reversibly between those of the slow-cooling and quenched regimes through electric/optical-pulse heating and subsequent rapid cooling. Such a non-volatile phase conversion is not predicated on the temperature-hysteresis that accompanies the first-order transition, thereby allowing for the switching operation over a wide temperature range.

## 2. Critical Cooling Rate

To kinetically avoid a first-order phase transition during cooling and reach the quenched regime, the cooling rate must exceed a threshold value, which we call the critical cooling rate, $R_c$. When considering the extent to which $R_c$ varies between materials in interacting electron/spin systems, it is instructive to refer to the accumulated knowledge for liquids. **Figure 2** summarizes the cooling-rate-dependent variation of low-temperature states in prototypical liquids;[1,10,11] here, we highlight two important aspects. First, as mentioned in the previous section, all simple liquids become crystals at low temperatures in the limit of slow cooling, whereas glass is realized in the limit of rapid cooling. Second, the critical cooling rate that separates the slow-cooling and quenched regimes substantially varies from liquid to liquid by more than 16 orders of magnitude. The wide variation in $R_c$ clearly indicates that the range of standard cooling rates ($2\times10^{-3}$–$4\times10^{-1}$ K/s) is often too narrow to explore both the slow-cooling and quenched regimes in a liquid of interest. For instance, molten $SiO_2$ always becomes glass within the standard cooling rates, but it can crystallize when much slower cooling rates are applied;[1] similarly, monoatomic metallic liquids, such as Ta, usually crystallize, but nevertheless, they can vitrify when an extremely high cooling rate, such as $10^{14}$ K/s, is achieved.[11] Thus, from Fig. 2, one can envisage that when pursuing the quenched regime in diverse interacting electron/spin systems, the broadest range of cooling rates possible must be explored.

To achieve cooling rates far above $10^{-1}$ K/s for condensed-matter systems, one can exploit rapid cooling following optical/electric-pulse heating of a sample anchored to a cryogenic system. This facile method enables cooling rates of $10^2$–$10^3$ K/s for bulk samples,[8,9] which are sufficiently high to reach the quenched regime, at least for some strongly correlated electron systems[5-8] and a non-centrosymmetric chiral magnet,[9] as described in the following sections.

## 3. Application of Rapid Cooling to Interacting Electron/Spin Systems

### 3.1. Charge-ordering System

To the best of our knowledge, only a few materials have $R_c$ values within the range of standard cooling rates, i.e., $2\times10^{-3}$–$4\times10^{-1}$ K/s. One such paradigmatic system is the organic conductor, $\theta$-(BEDT-TTF)$_2$RbZn(SCN)$_4$ (denoted as $\theta$-RbZn, where BEDT-TTF represents bis(ethylenedithio)tetrathiafulvalene).[12] The crystal structure is composed of alternating layers of conducting BEDT-TTF molecules and insulating monovalent RbZn(SCN)$_4$ anions, and the conduction band, which comprises the BEDT-TTF highest-occupied molecular orbital (HOMO), is hole-1/4-filled (that is, one hole per two BEDT-TTF molecules). **Figure 3**a illustrates the temperature- and cooling-rate-dependent electronic states in $\theta$-RbZn.[5] At high temperatures (above 200 K), the charges (holes) are well delocalized within the conducting BEDT-TTF layer; thus, the nominal valence is uniform and +0.5. This electronic state may therefore be described as liquid-like ("charge liquid"). When slowly cooled (< $2\times10^{-2}$ K/s), $\theta$-RbZn undergoes a first-order transition at ≈200 K, and below that, the charges segregate into hole-rich (+0.85) and hole-poor (+0.15) sites because of strong intersite Coulomb repulsion.[13-15] This periodic charge arrangement is accompanied by a sharp resistivity increase (Fig. 3b)[5,12] and is called a charge-ordered state or "charge crystal." Below, we show that, analogous to the case of classical liquids (Fig. 1b), the charge-ordering or charge-crystallization transition is kinetically avoided under rapid cooling (> $0.5–1\times10^{-1}$ K/s), resulting in a non-trivial charge glass as an electronic state of the quenched regime.[5]

The kinetic avoidance of the charge-ordering transition can be straightforwardly verified by measuring the temperature-resistivity profile (Fig. 3b).[5,12] Note that during rapid cooling (> $0.5–1\times10^{-1}$ K/s), the sharp resistivity increase is not observed, and the resistivity varies smoothly across 200 K (the charge-ordering temperature, $T_{CO}$, when slowly cooled). This continuous behavior demonstrates that the charge-liquid state persists below $T_{CO}$ as a supercooled state. Resistance fluctuation spectroscopy reveals that the charge fluctuations slow down as the temperature decreases and eventually freeze at the laboratory time scale; thus, the supercooled charge liquid takes on a frozen charge configuration, called charge glass.[5,6] Because the details of the charge glass are beyond the scope of this Research News contribution, here, we only enumerate the characteristics of the charge glass: (i) the valence of BEDT-TTF appears to be spatially inhomogeneous;[16] (ii) short-range orders or charge clusters, with distinct symmetry from that of the charge-ordered state, develop as the temperature decreases, but below a charge-glass transition temperature, the characteristic size of the charge clusters saturates at an intermediate value that is not long-range ordered;[5,6] (iii) the time scale of charge fluctuations is slower than the laboratory time scale,[5,6] as described; and (iv) aging behavior is observed in resistivity,[6] analogous to the dielectric aging in structural glass.[17]

Similar kinetic avoidance is also observed in another isostructural organic conductor, i.e., $\theta$-(BEDT-TTF)$_2$TlCo(SCN)$_4$ (denoted as $\theta$-TlCo), but only when an unconventionally high cooling rate, such as $10^3$ K/s, is applied.[7,8] By contrast, in yet another organic conductor, i.e., $\theta$-(BEDT-TTF)$_2$CsZn(SCN)$_4$ (denoted as $\theta$-CsZn), the charge liquid results in a charge glass, even under very slow cooling, such as $2\times10^{-3}$ K/s;[6] thus, we suppose that although the charge-ordering transition in $\theta$-CsZn has never been experimentally observed using slow cooling rates, $R_c$ exists and is even lower than $2\times10^{-3}$ K/s. We add the material dependence of $R_c$ to Fig. 2 as a reference.

The systematic variations in $R_c$ within the series of the isostructural compounds provide insights into which material parameter plays a key role in $R_c$ or, equivalently, the phase-transition speed. In considering the magnitude relation of $R_c$, we refer to the geometry of the BEDT-TTF anisotropic triangular lattice. As shown in Fig. 3a, the BEDT-TTF molecules are arranged to form a triangular network in the conducting layer; such a geometrically frustrated lattice potentially yields competition among various charge-ordering patterns,[18] as observed for antiferromagnetically interacting spins in a geometrically frustrated lattice. To characterize the degree of the geometrical frustration, here, we adopt the simplest parameter, $c/p$, i.e., the ratio of the two different triangular-lattice parameters (see the schematic in Fig. 3a).

When the phase diagrams are constructed according to $c/p$ and their cooling-rate dependence is considered (Figs. 3c-f),[5-8] we note two possible tendencies. First, as widely observed in frustrated spin systems, the transition temperature decreases as the frustration increases (that is, $c/p$ approaches unity), although $T_{CO}$ of $\theta$-CsZn is hypothetical. Second, the charge-ordered state in the more frustrated system is less robust against rapid cooling or, equivalently, has lower $R_c$. Given that a low $R_c$ is closely related to slow phase-transition kinetics, the observed tendency indicates that the more frustrated system tends to require more time to settle into the lowest-energy charge arrangements out of many competing patterns.[7] Thus, the tendencies observed in Figs. 3c-f imply that geometrical frustration generally affects not only the static aspects, such as transition temperature, but also the kinetic aspects of a phase transition, such as $R_c$.[7] These implications may also be relevant to frustrated spin systems; for instance, a frozen disordered spin state that sometimes emerges in frustrated magnets at the lowest temperature may result from the unintentional kinetic avoidance of some long-range order,[19] analogous to the case of $\theta$-CsZn.[6]

### 3.2. Topological Magnet

A first-order transition of interacting spin systems may be kinetically avoided if the utilized cooling rates are sufficiently high. A magnetic transition between topologically distinct, long-range-ordered spin textures is an ideal platform for exploring this issue because the difference in topology ensures a first-order-transition character between the two states. The non-centrosymmetric chiral magnet, MnSi, is an archetypal system that exhibits such a topological magnetic transition, from the topological magnetic skyrmion lattice (SkL) to the non-topological conical spin texture upon cooling.[20] Below, we show that this transition can be kinetically avoided under rapid cooling, thus yielding the metastable quenched SkL at low temperatures (**Figure 4**a),[9] as diamond grown in the mantle can persist even at room temperature as a metastable quenched state.

The magnetic skyrmion is a vortex-like, spin-swirling object, in which the local magnetic moments at the center and perimeter are antiparallel, and this spin texture is characterized by a nonzero topological winding number (or skyrmion number).[20-26] In a bulk crystal of MnSi, an aggregate of the magnetic skyrmions appear as a triangular lattice within the narrow temperature and magnetic-field region of the phase diagram (Fig. 4b).[20] The SkL is also characterized by nonzero winding number per magnetic unit cell; this topological nature exerts a so-called emergent magnetic field on the motion of conduction electrons through the coupling of the conduction-electron spin to the underlying spin texture.[27,28] As a result, a unique Hall effect (called the topological Hall effect) emerges in the SkL phase in addition to the normal and anomalous Hall effects;[29,30] thus, Hall resistivity, $\rho_{yx}$, can be exploited as a useful probe for studying the kinetic avoidance of the SkL-to-conical transition.

Figure 4d displays the magnetic-field profiles of $\rho_{yx}$ at 10 K after performing field-cooling at 0.22 T (along the <100> axis) with slow and rapid cooling rates.[9] When slowly cooled, $\rho_{yx}$ exhibits small values (on the order of 1 nΩ cm) in the considered magnetic-field range (0–1 T). In contrast, the $\rho_{yx}$ value at 0.22 T is dramatically enhanced after performing rapid field-cooling using an electric current pulse (up to ≈32 nΩ cm). When the magnetic field varies from 0.22 T, the $\rho_{yx}$ values remain enhanced in a certain magnetic-field range and then sharply drop to the values corresponding to the slow field-cooling case. The large enhancement of $\rho_{yx}$ is observed only when rapid field-cooling is performed to pass through the SkL phase and therefore indicates that by applying rapid cooling (≈700 K/s in this case: for details, see ref. [9]), the SkL-to-conical transition is kinetically avoided, and consequently, the SkL persists even at 10 K as a metastable quenched state.

In MnSi, the crossover behavior between the slow-cooling and quenched regimes has been studied in more detail. Figure 4e displays the cooling-rate dependence of $\rho_{yx}$ measured at 10 K and 0.22 T.[9] Note that the $\rho_{yx}$ value remains small in the slow-cooling regime (≈0–1 nΩ cm, corresponding to the non-topological spin-conical phase), but as the cooling rate becomes high, it steeply increases and then saturates above 100 K/s at an enhanced value (≈32 nΩ cm, corresponding to the quenched SkL). At intermediate cooling rates, $\rho_{yx}$ exhibits intermediate values because of imperfect kinetic avoidance of the SkL-to-conical transition, implying a phase mixture of the SkL and the conical state. If we define $R_c$ as the cooling rate at which $\rho_{yx}$ exhibits a half of the saturated value, $R_c$ is estimated to be ≈30 K/s, which is also added to Fig. 2 as a reference.

To gain more insight into the quenched SkL, it is helpful to see the phase diagram after performing rapid cooling at 0.22 T (Fig. 4c).[9] Remarkably, the quenched SkL region (or, more precisely, the region in which the quenched SkL has a prolonged lifetime) is found to be wide compared with the thermodynamically stable SkL phase. We also note that a narrow temperature gap of 2–3 K wide exists between the thermodynamically stable SkL phase and the quenched SkL region. The origin of this gap can be traced back to the shortened lifetime of the metastable quenched SkL in this gapped temperature region; in fact, although the lifetime (at 0.22 T) appears to exceed one year at 20 K, it is shortened to less than 10 seconds at 25-26 K, thereby allowing the metastable SkL to relax into the thermodynamically stable conical phase on the laboratory time scale.[9] This dramatic temperature dependence exemplifies the central concept in this Research News contribution, i.e., a temperature range in which a first-order transition can practically progress is limited, and therefore, a metastable quenched state with a prolonged lifetime can be realized when the temperature range is quickly traversed during rapid cooling.

## 4. Phase-change-memory (PCM) Function Implemented using Interacting Electron/Spin Systems

Once a metastable quenched charge/spin state is successfully obtained by implementing rapid cooling, a nonvolatile phase-change-memory (PCM) function that exploits thermodynamically stable and quenched states as rewritable state variables can be designed. Similar to the conventional PCM, which utilizes thermodynamically stable crystal and metastable glass as state variables,[31-33] the working principle of the nonvolatile PCM function implemented using interacting electron/spin systems is predicated on manipulating the lifetime of the metastable quenched state; namely, whereas the quenched state is practically stable at low temperatures because of its prolonged lifetime, the lifetime is dramatically shortened when the temperature is elevated, facilitating the relaxation from the metastable quenched state to the thermodynamically most-stable state.

We explain the operation protocol of the PCM function by referring to two cases: the charge-ordering system (**Figure 5**a)[8] and the skyrmion system (Fig. 5d).[9] Starting with the thermodynamically stable state of the slow-cooling regime, such as the charge-ordered state (or the spin-conical state), an electric/optical single pulse is applied to heat the sample, which is anchored to a cryogenic heat bath, to a temperature above the first-order transition; thus, the charge liquid (or SkL) is realized during the pulse. After the pulse ceases, rapid cooling starts, and the charge-ordering transition (or SkL-to-conical transition) is kinetically avoided, resulting in a metastable charge/spin state of the quenched regime. In this way, the phase conversion from the charge-ordered (or spin-conical) state to the charge glass (or the quenched SkL) is completed ("SET" process). To revert to the state of the slow-cooling regime, a pulse with weaker intensity and longer pulse duration is used to heat the sample to a temperature slightly below the first-order transition, at which a shortened lifetime of the metastable quenched state facilitates the relaxation to the thermodynamically stable state (here, the pulse width should be tuned so that the phase conversion can be completed within the duration). After the pulse ends, the sample temperature quickly returns to the heat-bath

WILEY-VCHtemperature without affecting the thermodynamically stable state realized; accordingly, the original charge-ordered (or spin-conical) state is retrieved ("RESET" process). We note that during the SET and RESET processes, the heat-bath temperature and/or magnetic fields do not need to be purposely changed. We also note that unlike the conventional PCM based on atomic crystal/glass, the PCM function described here does not require explicit melting of the background lattice of atoms/molecules.

Figure 5b shows the time profiles during the single-cycle operation of the charge PCM, implemented using $\theta$-TlCo at 100 K,[8] which is well below the $T_{CO}$ of $\theta$-TlCo (≈245 K). By applying rectangular voltage pulses to quickly control the sample temperature (Fig. 5b, bottom panel), the electronic state is converted between the high-resistivity charge-ordered state and the low-resistivity charge-glass state (Fig. 5b, upper panel).[8] This operation is repeatable (Fig. 5c),[8] highlighting the deterministic nature of the working principle described. Similarly, the SkL and the conical states are electrically interconverted (Figs. 5e,f);[9] here, we show the spin PCM function operated at 10 K, well below the SkL-to-conical transition temperature (≈27 K). By applying electric pulses to achieve rapid cooling (SET process) or to heat the MnSi crystal to the gapped temperature region in the phase diagram (Fig. 4c) (RESET process), the $\rho_{yx}$ value reproducibly switches between large and small values, signaling the creation and annihilation of the metastable SkL, respectively.[9]

In previously published studies, non-volatile switching between thermodynamically stable and metastable states has often been demonstrated at temperatures within the hysteresis region that accompanies a thermally driven first-order transition.[34,35] We note that this conventional approach inevitably puts constraints on the operation temperature; that is, a material incorporated in a memory device should be kept within the temperature range of the hysteresis. By contrast, the approach that exploits a quenched state can function outside the hysteresis region, allowing operation over a much wider temperature range; therefore, the design principle highlighted in this Research News contribution is beneficial, particularly for first-order-transition systems with little temperature hysteresis.

## 5. Conclusion and Outlook

We have demonstrated, at least for some materials, that interacting electron/spin systems undergoing a first-order transition show either a thermodynamically stable state or a metastable quenched state at low temperatures, depending on the cooling rate. To realize electronic/magnetic states in the quenched regime, experimental cooling rates should exceed a critical value, which is strongly material dependent. Moreover, the critical cooling rate is currently difficult to predict because of the lack of microscopic knowledge about, for instance, the interface structure of the nucleus in individual materials; a kinetic nucleation pathway or Ostwald's step rule may be another issue to be considered.[36-38] Therefore, in pursuing the quenched regime in a material of interest, high cooling rates should be attempted to the greatest possible extent. Because we have succeeded in quenching the charge and spin degrees of freedom in condensed matter, it seems reasonable that another degree of freedom, i.e., orbital, could be quenched as well. We also expect that a quenched state potentially exhibits further symmetry breaking towards another metastable state. Such a metastable state derived from a quenched state may be described as a hidden state, which is not found in the thermo-equilibrium phase diagram.

In the context of the possible applications of quenched electronic/magnetic states, we designed and demonstrated non-volatile phase-change-memory functions. Although the operation speed remains fairly slow and the RESET process requires ≈10 s, identifying a material exhibiting faster phase-transition kinetics would greatly improve these issues. Such materials likely have critical cooling rates far above $10^2$–$10^3$ K/s; thus, to develop faster memory function, a quenching method must also be developed. Local heat management using a focused nanosecond/picosecond-pulse laser facilitates achieving high cooling rates that exceed $10^6$–$10^7$ K/s and will be useful in quenching high-speed phase-transition materials.

As highlighted, the application of rapid cooling can dramatically alter the low-temperature states of interacting electron/spin systems beyond the framework of the thermo-equilibrium phase diagram. This emerging phase-control method may open up a new avenue for basic research and applications of interacting electron/spin systems.

**Acknowledgements**
F.K. and H.O. thank H. Tanaka and T. Sato for their fruitful discussions. This work was partially supported by JSPS KAKENHI (Grant Nos. 25220709 and 15H05459).

Received: April 13, 2016
Revised: May 5, 2016
Published online: 21 June 2016

[1]    D. R. Uhlmann, *J. Non-Cryst. Solids* **1972**, 7, 337–348.

[2]    C. A. Angell, *Science* **1995**, 267, 1924–1935.

[3]    P. G. Debenedetti, F. H. Stillinger, *Nature* **2001**, 410, 259–267.

[4]    F. P. Bundy, W. A. Bassett, M. S. Weathers, R. J. Hemley, H. U. Mao, A. F. Goncharov, *Carbon* **1996**, 34, 141–153.

[5]    F. Kagawa, T. Sato, K. Miyagawa, K. Kanoda, Y. Tokura, K. Kobayashi, R. Kumai, Y. Tokura, *Nat. Phys.* **2013**, 9, 419–422.

[6]    T. Sato, F. Kagawa, K. Kobayashi, K. Miyagawa, K. Kanoda, R. Kumai, Y. Murakami, Y. Tokura, *Phys. Rev. B* **2014**, 89, 121102.

[7]    T. Sato, F. Kagawa, K. Kobayashi, A. Ueda, H. Mori, K. Miyagawa, K. Kanoda, R. Kumai, Y. Murakami, Y. Tokura, *J. Phys. Soc. Jpn.* **2014**, 83, 083602.

[8]    H. Oike, F. Kagawa, N. Ogawa, A. Ueda, H. Mori, M. Kawasaki, Y. Tokura, *Phys. Rev. B* **2015**, 91, 041101.

[9]     H. Oike, A. Kikkawa, N. Kanazawa, Y. Taguchi, M. Kawasaki, Y. Tokura, F. Kagawa, *Nat.Phys.* **2016**, 12, 62–66.

[10]    Z. Strnad, *Glass-Ceramic Materials*, Glass Science and Technology Vol. 8, Elsevier, Amsterdam.

[11]    L. Zhong, J. Wang, H. Sheng, Z. Zhang, S. X. Mao, *Nature* **2014**, 512, 177–180.

[12]    H. Mori, S. Tanaka, T. Mori, *Phys. Rev. B* **1998**, 57, 12023–12029.

[13]    M. Watanabe, Y. Noda, Y. Nogami, H. Mori, *J. Phys. Soc. Jpn.* **2004**, 73, 116–122.

[14]    K. Miyagawa, A. Kawamoto, K. Kanoda, *Phys. Rev. B* **2000**, 62, 7679–7682.

[15]    K. Yamamoto, K. Yakushi, K. Miyagawa, K. Kanoda, A. Kawamoto, *Phys. Rev. B* **2002**, 65, 085110.

[16]    K. Suzuki, K. Yamamoto, K. Yakushi, A. Kawamoto, *J. Phys. Soc. Jpn.* **2005**, 74, 2631–2639.

[17]    R. L. Leheny, S. R. Nagel, *Phys. Rev. B* **1998**, 57, 5154–5162.

[18]    H. Seo, *J. Phys. Soc. Jpn.* **2000**, 69, 805–820.

[19]    D. Pomaranski, L. R. Yaraskavitch, S. Meng, K. A. Ross, H. M. L. Noad, H. A. Dabkowska, B. D. Gaulin, J. B. Kycia, *Nat. Phys.* **2013**, 9, 353–356.

[20]    S. Mühlbauer, B. Binz, F. Jonietz, C. Pfleiderer, A. Rosch, A. Neubauer, R. Georgii, P. Böni, *Science* **2009**, 323, 915–919.

[21]    A. Bogdanov, D. A. Yablonskii, *Sov. Phys. JETP* **1989**, 68, 101–103.

[22]    A. Bogdanov, A. Hubert, *J. Magn. Magn. Mater.* **1994**, 138, 255–269.

[23]    X. Z. Yu, Y. Onose, N. Kanazawa, J. H. Park, J. H. Han, Y. Matsui, N. Nagaosa, Y. Tokura *Nature* **2010**, 465, 901–904.

[24]    N. Nagaosa, Y. Tokura, *Nat. Nanotech.* **2013**, 8, 899–911.

[25]    H. B. Braun, *Adv. Phys.* **2012**, 61, 1–116.

[26]    Y. Zhou, M. Ezawa, *Nat. Commun.* **2014**, 5, 4652.

[27] J. Ye, Y. B. Kim, A. J. Millis, B. I. Shraiman, P. Majumdar, and Z. Tešanović, *Phys. Rev. Lett.* **1999**, 83, 3737–3740.

[28] P. Bruno, V. K. Dugaev, M. Taillefumier, *Phys. Rev. Lett.* **2004**, 93, 096806.

[29] M. Lee, W. Kang, Y. Onose, Y. Tokura, N. P. Ong, *Phys. Rev. Lett.* **2009**, 102, 186601.

[30] A. Neubauer, C. Pfleiderer, B. Binz, A. Rosch, R. Ritz, P. G. Niklowitz, P. Böni, *Phys. Rev. Lett.* **2009**, 102, 186602.

[31] S. R. Ovshinsky, *Phys. Rev. Lett.* **1968**, 21, 1450–1453.

[32] M. Wuttig, N. Yamada, *Nat. Mat.* **2007**, 6, 824–832.

[33] S. Raoux, *Annu. Rev. Mater. Res.* **2009**, 39, 25–48.

[34] S. Koshihara, Y. Tokura, K. Takeda, T. Koda, *Phys. Rev. Lett.* **1992**, 68, 1148–1151.

[35] N. Takubo, Y. Ogimoto, M. Nakamura, H. Tamaru, M. Izumi, K. Miyano, *Phys. Rev. Lett.* **2005**, 95, 017404.

[36] W. Z. Ostwald, *Physik. Chem.* **1897**, 22, 289–330.

[37] Y. Peng, F. Wang, Z. Wang, A. M. Alsayed, Z. Zhang, A. G. Yodh, Y. Han, *Nat. Mat.* **2015**, 14, 101–108.

[38] J. Russo, F. Romano, H. Tanaka, *Nat. Mat.* **2014**, 13, 733–739.

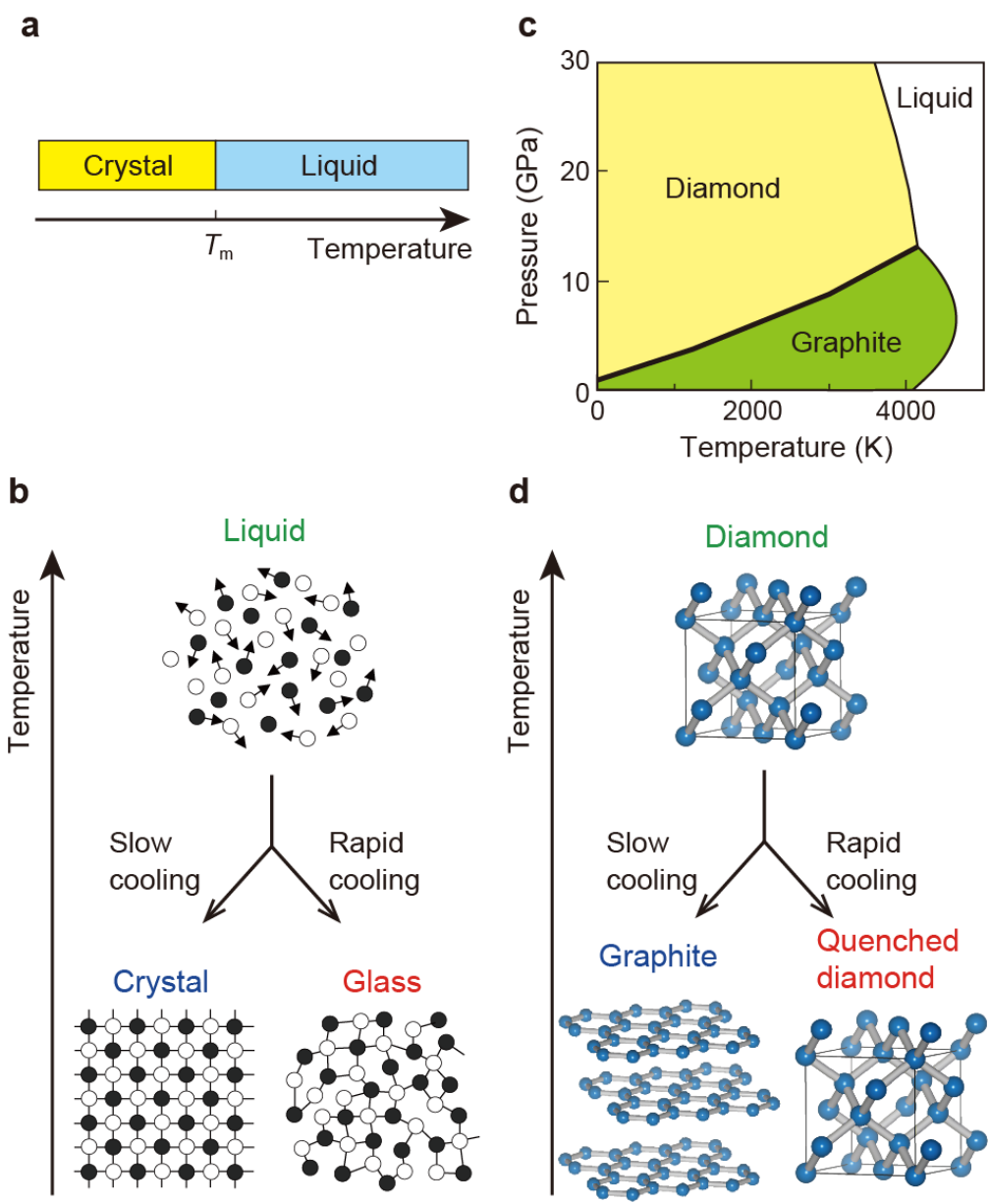

**Figure 1.** Competition between a thermodynamically stable state and a metastable quenched state. a,c) The thermo-equilibrium phase diagram of liquid (a) and aggregation of carbon atoms (c). b,d) Schematic of the cooling-rate-dependent bifurcation of liquid (b) and diamond (d) during cooling.

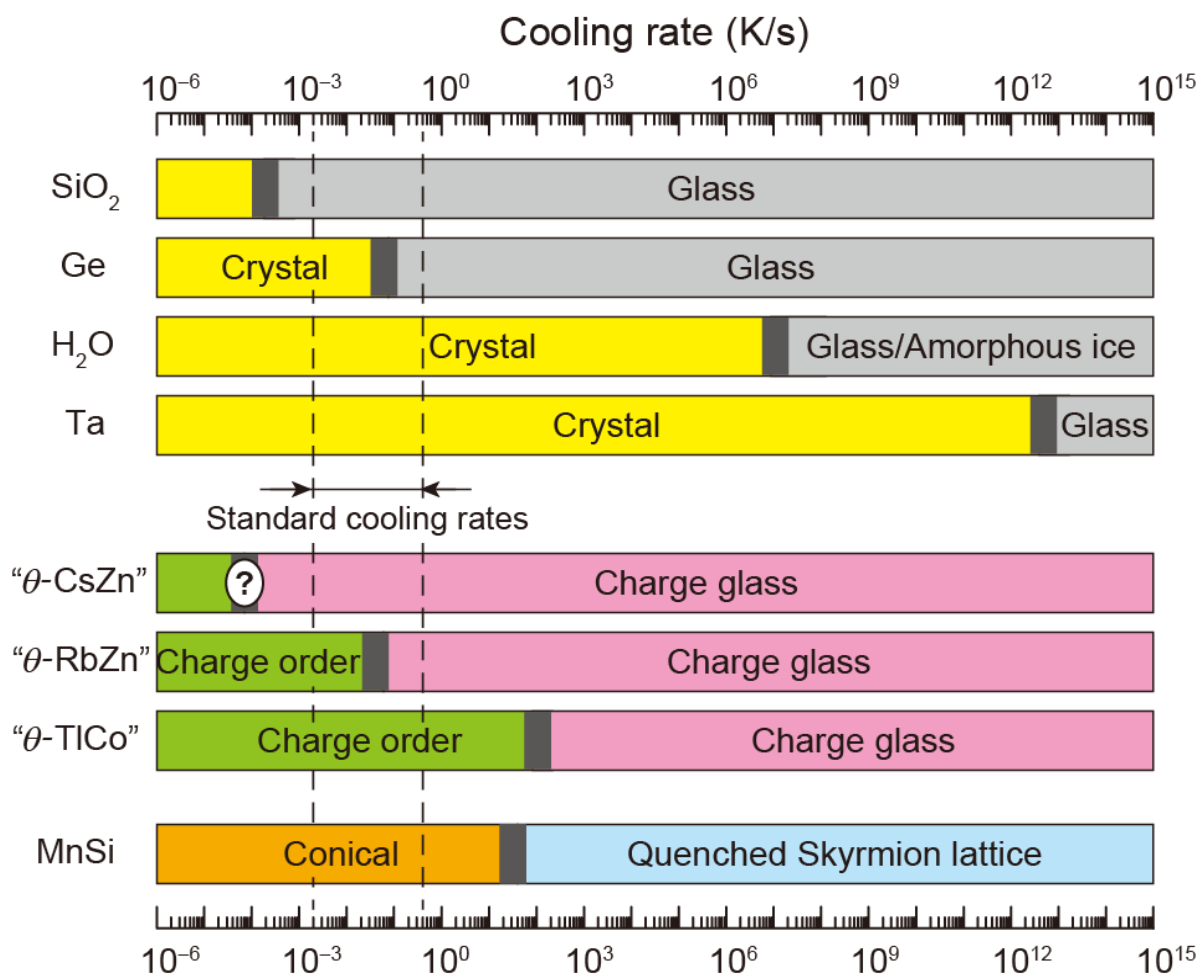

**Figure 2.** Enumeration of the cooling-rate-dependent variation of the low-temperature state. Prototypical liquids and the interacting electron/spin systems highlighted in this Research News contribution are listed from previously published studies.[1,5-11]. $\theta$-$MM'$ ($MM'$ = CsZn, RbZn, or TlCo) represents the organic conductor, i.e., $\theta$-(BEDT-TTF)$_2MM'$(SCN)$_4$. The boundaries separating the slow-cooling and quenched regimes represent critical cooling rates. Because the transition between the two regimes is crossover,[1,9] the boundaries are depicted with broad lines. The charge order in $\theta$-CsZn has experimentally never been observed using slow cooling rates (down to $\approx 2 \times 10^{-3}$ K/s);[6] thus, the critical cooling rate of $\theta$-CsZn is hypothetical.

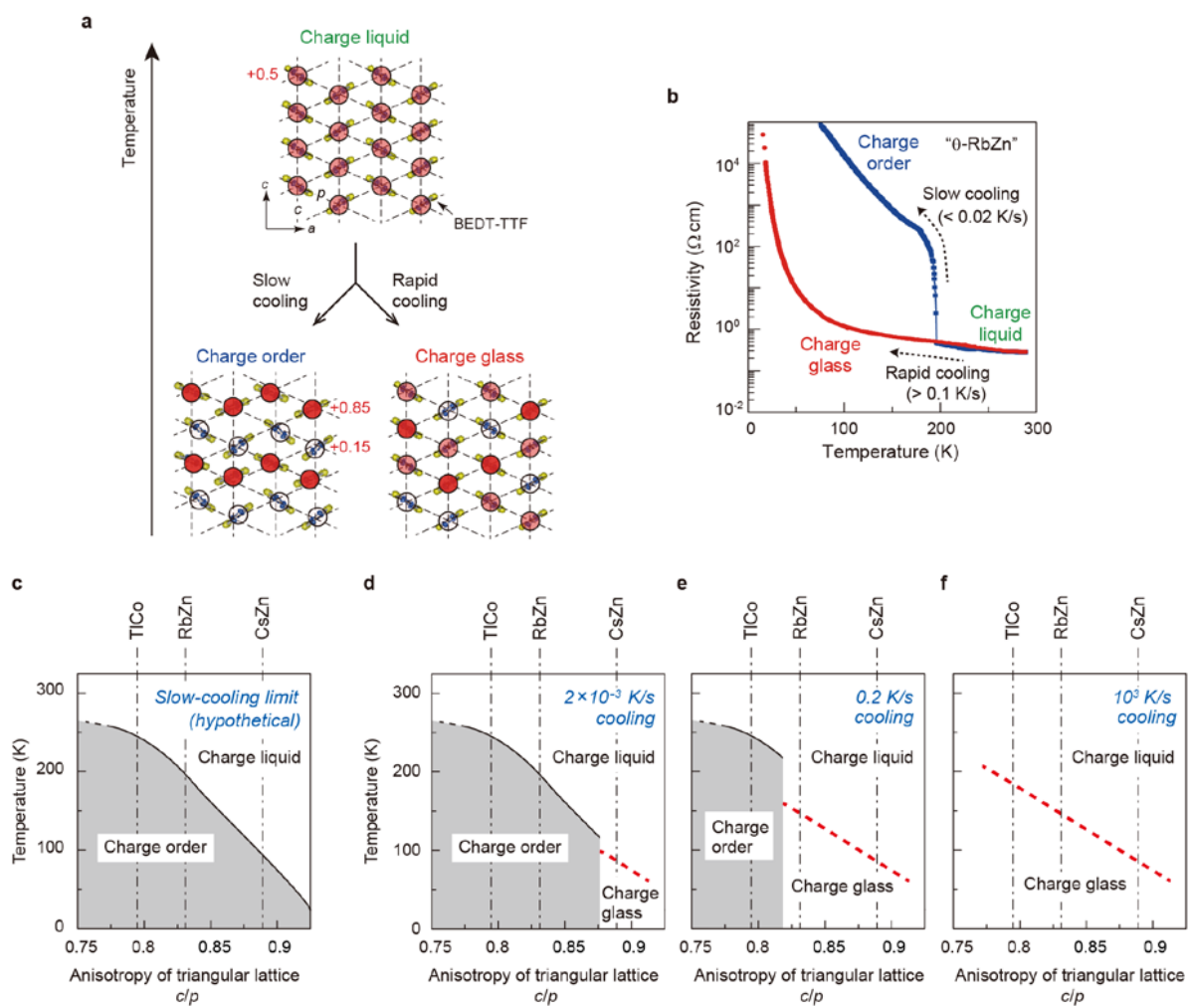

**Figure 3.** Application of rapid cooling to the organic conductors $\theta$-(BEDT-TTF)$_2$$MM'$(SCN)$_4$ ($MM'$ = CsZn, RbZn, or TlCo). a) Schematic of the cooling-rate-dependent bifurcation of the charge liquid at high temperatures into the charge order and the charge glass at low temperatures. b) The temperature dependence of the resistivity during cooling on different cooling rates in $\theta$-(BEDT-TTF)$_2$RbZn(SCN)$_4$ (denoted as $\theta$-RbZn). c-f) Cooling-rate dependence of the phase diagram of $\theta$-(BEDT-TTF)$_2$$MM'$(SCN)$_4$, plotted with respect to the ratio of the two different triangular lattice parameters, $c/p$. The definitions of the lattice parameters $c$ and $p$ are illustrated in (a) (see the schematic of the charge liquid). (c) represents the hypothetical phase diagram corresponding to the slow-cooling limit. The figure panels are constructed on the basis of ref. [5-8].

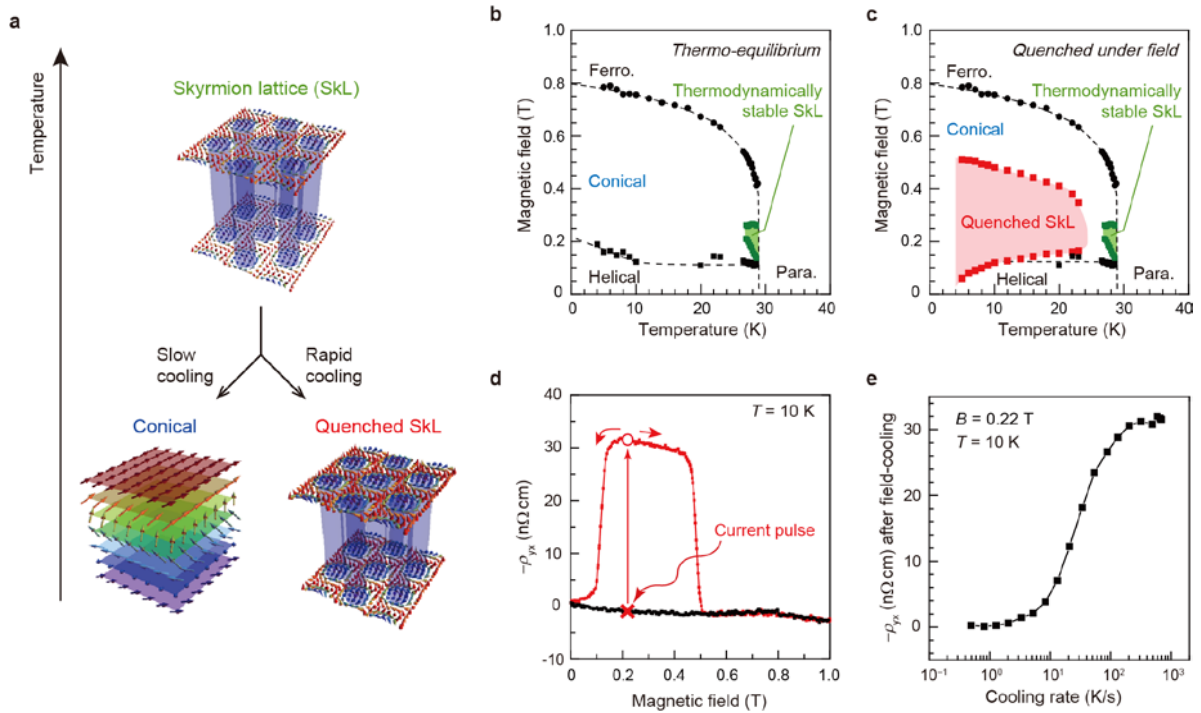

**Figure 4.** Application of rapid cooling to the topological magnet MnSi. a) Schematic of the cooling-rate-dependent bifurcation of the thermodynamically stable skyrmion lattice (SkL) at high temperatures into the conical order and quenched SkL at low temperatures. b,c) Magnetic phase diagrams under thermo-equilibrium (b) and quenched conditions (c). d) Magnetic-field dependence of the Hall resistivity, $\rho_{yx}$, at 10 K, measured before and after quenching at 0.22 T. By applying a current pulse to the equilibrium state under a magnetic field of 0.22 T (the cross), the $\rho_{yx}$ value becomes enhanced (the open circle). e) Cooling-rate dependence of $\rho_{yx}$ at 10 K and 0.22 T after field-cooling with different cooling rates. The figure panels are constructed on the basis of ref. [9].

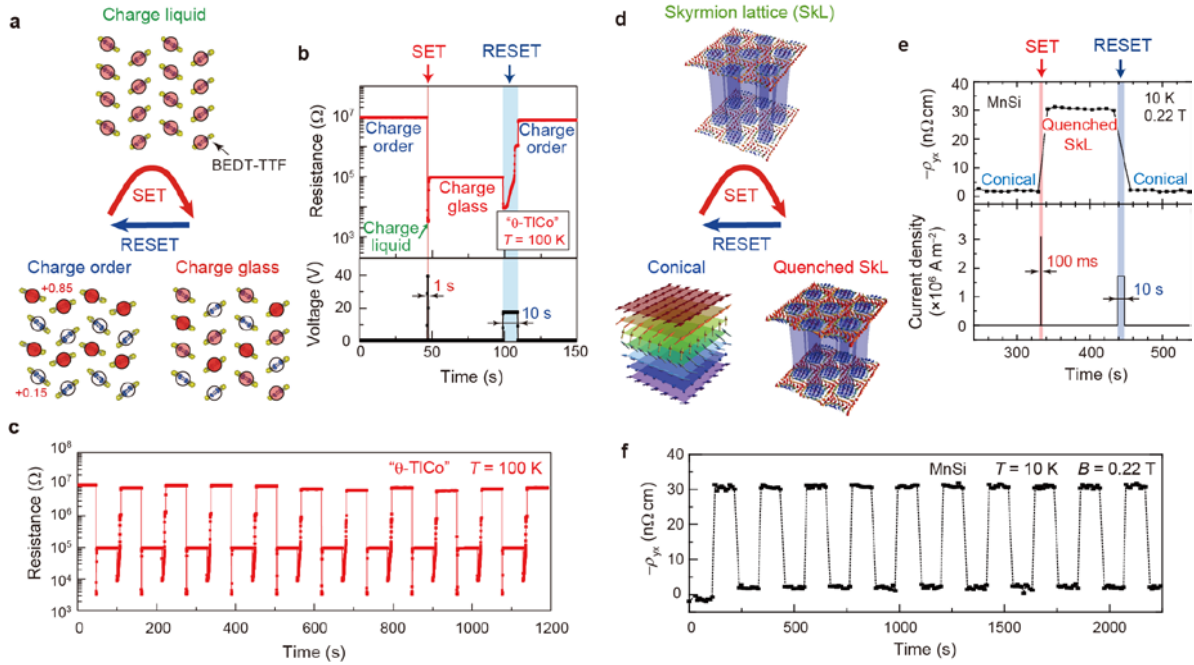

**Figure 5.** Phase-change-memory (PCM) functions implemented using interacting electron/spin systems. a) A scheme for the rewritable switching of the charge PCM implemented using $\theta$-(BEDT-TTF)$_2$TlCo(SCN)$_4$ (denoted as $\theta$-TlCo). b) Single-cycle operation of the SET and RESET processes under the application of rectangular voltage pulses. The upper and bottom panels present the time profiles of the two-probe resistance connected in series to a load resistor of 30 k$\Omega$ and the voltage applied to the circuit, respectively. c) Repetitive switching between the charge glass and the thermodynamically stable charge-ordered phase. d) A scheme for the rewritable switching of the spin PCM implemented using MnSi. e) Single-cycle operation of the SET and RESET processes under the application of rectangular current pulses. The upper and bottom panels display the time profiles of $\rho_{yx}$ and the applied current density, respectively. f) Repetitive switching between the quenched SkL and the thermodynamically stable conical phase. The figure panels of the charge PCM and spin PCM are constructed on the basis of ref [8,9], respectively.

21